\documentclass[preprint,12pt]{elsarticle}

\usepackage{amssymb}

\journal{Journal of Computational Physics}

\begin{document}

\begin{frontmatter}



\title{A conservative scheme for the relativistic Vlasov-Maxwell system}


\author{Akihiro Suzuki}
\author{Toshikazu Shigeyama}

\address{Research Center for the Early Universe, School of Science, University of Tokyo, Hongo, 7-3-1, Bunkyo-ku, Tokyo, Japan}

\begin{abstract} 
A new scheme for numerical integration of the 1D2V relativistic Vlasov-Maxwell system is proposed. 
Assuming that all particles in a cell of the phase space move with the same velocity as that of the particle located at the center of the cell at the beginning of each time step, we successfully integrate the system with no artificial loss of particles.
Furthermore, splitting the equations into advection and interaction parts, the method conserves the sum of the kinetic energy of particles and the electromagnetic energy. 
Three test problems, the gyration of particles, the Weibel instability, and the wakefield acceleration, are solved by using our scheme. 
We confirm that our scheme can reproduce analytical results of the problems. 
Though we deal with the 1D2V relativistic Vlasov-Maxwell system, our method can be applied to the 2D3V and 3D3V cases. 
\end{abstract}


\end{frontmatter}



\section{Introduction}
In a wide range of plasma processes operating in laboratories or astrophysical phenomena, interactions between relativistic particles and electromagnetic fields play vital roles. 
For instance, recent laser experiments revealed that a high intensity laser can accelerate particles to ultra-relativistic speed (see, e.g., \citet{e96}). 
Non-thermal components found in spectra of active astrophysical objects, e.g., supernova remnants, gamma-ray bursts, and jets from active galactic nuclei, are interpreted as synchrotron radiation emitted by charged, relativistic particles gyrating about magnetic field lines. 

There are two approaches in order to model such plasmas. 
One is the fluid approach based on the relativistic magnetohydrodynamics(RMHD) and the other is the kinetic approach based on the Boltzmann equation coupled with the Maxwell equations. 
Because the fluid approach implicitly assumes that the distribution of particles in the momentum space is the Maxwell-J\"uttner distribution, which is the relativistic extension of the classical Maxwell-Boltzmann distribution, the kinetic approach is indispensable for dealing with the momentum distribution deviating from the thermal equilibrium.  
Especially, dilute plasmas in which collisions between particles composing the plasmas are absent, often called collisionless plasmas, are known to be modeled by the so-called Vlasov-Maxwell system \cite{s94}. 
  
At present, the most reliable and reasonable method to simulate dynamical behaviors of collisionless plasmas is the particle-in-cell (PIC) simulation (see, e.g., \citet{BL91}), which calculates the orbits of charged particles by solving the equation of motion and the configuration of electromagnetic fields by solving the Maxwell equations. 
In this method, the momentum distribution of plasmas is approximated by an ensemble of the momentum of each particle placed in the physical space. 
Although the number of particles in virtual plasmas produced by a PIC simulation is much smaller than that in real plasmas, it is known that behaviors of plasmas are well reproduced by the method. 
Nevertheless, we cannot avoid significant numerical noises due to the shortage of particles when we focus on the high momentum tail of the distribution function of plasma particles. 

On the other hand, the direct numerical integration of the Vlasov-Maxwell system (referred to as "Vlasov simulation"), which discretizes the momentum space as well as the physical space, does not suffer from such noises. 
Therefore, some methods to perform the Vlasov simulation have been developed \cite{ck76,f99,ny99,mcct02}. 
While Vlasov simulations require higher computational performance than PIC simulations do, recent developments of computational technology allow us to study plasma processes by using Vlasov simulations. 
For example, \citet{mcct02} presented a scheme to numerically integrate the 2D3V Vlasov-Maxwell system (the term "2D3V" means the two dimensional space associated with the three dimensional velocity space) and demonstrated that the scheme could simulate the Weibel instability with high accuracy. 
\citet{vvm05} provided a scheme for the integration of the electrostatic 1D2V Vlasov-Poisson system in a uniform magnetic field. 
They adopted the polar coordinates in the velocity space, which allows them to perform simulations with a good energy conservation. 
The scheme presented by \citet{vtchm07} integrates the Vlasov-Maxwell system in the hybrid approximation, i.e., they solve the 2D3V electromagnetic Vlasov equation for ions and fluid equations for electrons, based on the current advance method introduced by \citet{matthews}.  
\citet{SS09} investigated non-linear behavior of the Weibel instability in detail by using a scheme similar to that of \citet{mcct02}. 
\citet{SG06} performed a series of simulations for the magnetic reconnection and confirmed that their results are consistent with those of some PIC simulations. 
However, the attempts stated above treated only non-relativistic plasmas. 
Investigations into the numerical integration of the relativistic Vlasov-Maxwell system are still rare. 
Although \citet{blgsb08} presented a scheme for the 1D2V relativistic Vlasov-Maxwell system, they assumed that particles have no dispersion in the transverse momentum space. 
Furthermore, there exists another problem that the mass and energy conservations are not always guaranteed unlike PIC simulations. 

In this paper, we propose a new conservative scheme for the numerical integration of the 1D2V relativistic Vlasov-Maxwell system that allows particles to have dispersions in the momentum space. 
The scheme is based on the semi-Lagrangian approach, which is extensively used to solve the Vlasov-Maxwell system \cite{blgsb08,s96,bs03}. 
In Sec. \ref{5:formulation}, we introduce the 1D2V relativistic Vlasov-Maxwell system and some characteristic scales, and then transform the equations for convenience of the subsequent sections. 
Sec. \ref{5:scheme} describes the method for the numerical integration of the system. 
In Sec. \ref{5:test}, we calculate three test problems using the scheme proposed in Sec. \ref{5:scheme}, 
the gyration of particles, the Weibel instability, and the wakefield acceleration. 
We conclude this paper in Sec. \ref{5:conclusion}.

\section{Formulation\label{5:formulation}}
In this section, we present a scheme for the numerical integration of the 1D2V relativistic Vlasov-Maxwell system. 
\subsection{The relativistic Vlasov-Maxwell system}
We consider a plasma whose spatial distribution varies along one direction, which implies that only two components of the momenta of particles, the longitudinal and the lateral components, need to be calculated. 
Thus, the 1D2V Vlasov equation for species $s$ describes the kinetic evolution of the distribution function $f^s(x,p,q,t)$ ($s=\mathrm{e}$ for electrons and $s=\mathrm{i}$ for ions) in the phase space $(x,p,q)$, where $x$ is the coordinate in the physical space, $p$ is the corresponding coordinate in the momentum space, and $q$ is the coordinate labeling the lateral momentum. 
In this case, the relativistic Vlasov equation takes the following form,
\begin{equation}
\frac{\partial f^s}{\partial t}+\frac{p}{m_s\Gamma^s}\frac{\partial f^s}{\partial x}
+Q_s\left(E^\parallel+\frac{q}{m_sc\Gamma^s}B^\perp\right)\frac{\partial f^s}{\partial p}
+Q_s\left(E^\perp-\frac{p}{m_sc\Gamma^s}B^\perp\right)\frac{\partial f^s}{\partial q}=0,
\label{5:1}
\end{equation}
where
\begin{equation}
\Gamma^s=\sqrt{1+\left(\frac{p}{m_sc}\right)^2+\left(\frac{q}{m_sc}\right)^2}
\end{equation}
represents the Lorentz factor. 
The constants $m_s$ and $Q_s$ represent the mass and the charge of a species $s$. 
$c$ is the speed of light. 
The electric field appearing here has two components parallel $E^\parallel$ and normal $E^\perp$ to the $x$-axis, while the magnetic field has only one component $B^\perp$ normal to the $x$-axis. 
Here the normal component of the electric field points to the direction of the lateral momentum and the electric and magnetic fields are perpendicular to each other. 
Thus, they are expressed as vector forms ${\bf E}=(E^\parallel,E^\perp,0)$ and ${\bf B}=(0,0,B^\perp)$ when the momentum vector is expressed as ${\bf p}=(p,q,0)$. 
Their time evolutions are governed by the Maxwell equations,
\begin{equation}
\frac{\partial E^\parallel}{\partial t}=-4\pi J^\parallel,\ \ \ 
\frac{1}{c}\frac{\partial E^\perp}{\partial t}+\frac{\partial B^\perp}{\partial x}=-4\pi J^\perp,\ \ \ 
\frac{1}{c}\frac{\partial B^\perp}{\partial t}+\frac{\partial E^\perp}{\partial x}=0,\ \ \ 
\end{equation}
where the electric current densities $J^\parallel$ and $J^\perp$ are expressed in terms of $f^s(x,p,q,t)$ as
\begin{eqnarray}
J^\parallel&=&\sum_s Q_s\int^\infty_{-\infty}\int^\infty_{-\infty}\frac{p}{m_sc\Gamma^s}f^s(x,p,q,t)dpdq\\
J^\perp&=&\sum_s Q_s\int^\infty_{-\infty}\int^\infty_{-\infty}\frac{q}{m_sc\Gamma^s}f^s(x,p,q,t)dpdq.
\end{eqnarray}

\subsection{Normalization}
For the numerical integration of the equations introduced above, we define the characteristic value for each physical quantity: $1/\omega_\mathrm{e}$ as the time scale, $c/\omega_\mathrm{e}$ as the length scale, $m_\mathrm{e}c$ as the momentum, $cm_\mathrm{e}\omega_\mathrm{e}/e$ as the electromagnetic field, and $m_e\omega_e^2/(4\pi e)$ as the electric current density. 
Here $\omega_\mathrm{e}$ is the electron plasma frequency defined by
\begin{equation}
\omega_\mathrm{e}^2=\frac{4\pi e^2n_0}{m_\mathrm{e}},
\end{equation}
where $e$ is the elementary charge and $n_0$ is the number density. 
Normalizing physical variables with these quantities and using the same notations for normalized quantities, one can obtain the dimensionless Vlasov equation,
\begin{equation}
\frac{\partial f^s}{\partial t}+\frac{p}{\Gamma^s}\frac{\partial f^s}{\partial x}
+R_q^s\left[E^\parallel+\frac{q}{\Gamma^s}B^\perp\right]\frac{\partial f^s}{\partial p}
+R_q^s\left[E^\perp-\frac{p}{\Gamma^s}B^\perp\right]\frac{\partial f^s}{\partial q}=0,
\label{5:vlasov}
\end{equation}
where the Lorentz factor is modified to
\begin{equation}
\Gamma^s=\sqrt{(R_m^s)^2+p^2+q^2}.
\label{mLorentz}
\end{equation}
Here $R_m^s$ and $R_q^s$ are dimensionless constants defined by
\begin{equation}
R_m^s=\frac{m_s}{m_e},\ \ \ 
R_q^s=\frac{Q_s}{e},
\end{equation}
respectively.
On the other hand, the Maxwell equations lead to the following dimensionless form,
\begin{equation}
\frac{\partial E^\parallel}{\partial t}=-J^\parallel,\ \ \ 
\frac{\partial E^\perp}{\partial t}+\frac{\partial B^\perp}{\partial x}=-J^\perp,\ \ \ 
\frac{\partial B^\perp}{\partial t}+\frac{\partial E^\perp}{\partial x}=0,\ \ \ 
\label{5:maxwell}
\end{equation}
where the dimensionless electric current densities $J^\parallel$ and $J^\perp$ are expressed in terms of $f^s(x,p,q,t)$ as
\begin{eqnarray}
J^\parallel&=&\sum_s R_q^s\int^\infty_{-\infty}\int^\infty_{-\infty}\frac{p}{\Gamma^s}f^s(x,p,q,t)dpdq,\\
J^\perp&=&\sum_s R_q^s\int^\infty_{-\infty}\int^\infty_{-\infty}\frac{q}{\Gamma^s}f^s(x,p,q,t)dpdq.
\label{5:current}
\end{eqnarray}
These two relations close the system.

\subsection{Transformation of equations}
For convenience of the following sections, we transform equation (\ref{5:vlasov}) into the conservative form and equation (\ref{5:maxwell}) into the advection form. 

Multiplying equation (\ref{5:vlasov}) by $\Gamma^s(p,q)$ and some algebraic manipulations lead to
the following equation,
\begin{eqnarray}
\frac{\partial (\Gamma^sf^s)}{\partial t}+\frac{p}{\Gamma^s}\frac{\partial (\Gamma^s f^s)}{\partial x}
+R_q^s\left(E^\parallel+\frac{q}{\Gamma^s}B^\perp\right)\frac{\partial (\Gamma^sf^s)}{\partial p}&&
\nonumber\\
+R_q^s\left(E^\perp-\frac{p}{\Gamma^s}B^\perp\right)\frac{\partial (\Gamma^sf^s)}{\partial q}
=R_q^s\frac{pE^\parallel+qE^\perp}{\Gamma^s}f^s.&&
\label{5:energy}
\end{eqnarray}
The L.H.S of this equation represents the advection of the rest and kinetic energy of particles along the characteristics of the Vlasov equation (\ref{5:vlasov}), while the R.H.S is interpreted as the exchange of energy between particles and electromagnetic fields.

On the other hand, introducing the following variables,
\begin{equation}
G(x)=\frac{E^\perp(x)+B^\perp(x)}{2},\ \ \ 
H(x)=\frac{E^\perp(x)-B^\perp(x)}{2},
\end{equation}
one can rewrite the Maxwell equations for the perpendicular components $E^\perp$ and $B^\perp$ as
\begin{equation}
\frac{\partial G}{\partial t}+\frac{\partial G}{\partial x}=-\frac{J^\perp}{2},\ \ \ 
\frac{\partial H}{\partial t}-\frac{\partial H}{\partial x}=-\frac{J^\perp}{2}.
\end{equation}
In the following, we integrate the above equations instead of the equations for the components $E^\perp$ and $B^\perp$.

\section{Strategy for numerical integration\label{5:scheme}}
In this section, we describe a method for numerical integration of the dimensionless Vlasov-Maxwell system (\ref{5:vlasov})-(\ref{5:energy}) introduced in the previous section. 

\subsection{Discretization}
First, we divide the phase space with the range of $[x_\mathrm{min},x_\mathrm{max}]\times[p_\mathrm{min},p_\mathrm{max}]\times[q_\mathrm{min},q_\mathrm{max}]$ into $N_x\times N_p\times N_q$ small cells each of which has the volume $\Delta x\Delta p\Delta q$. 
Thus, $\Delta x$, $\Delta p$, and $\Delta q$ are 
\begin{equation}
\Delta x=\frac{x_\mathrm{max}-x_\mathrm{min}}{N_x},\ \ \ 
\Delta p=\frac{p_\mathrm{max}-p_\mathrm{min}}{N_p},\ \ \ 
\Delta q=\frac{q_\mathrm{max}-q_\mathrm{min}}{N_q}. 
\end{equation}
The center of a cell labeled by integers $(i,j,k)$ is located at $(x,p,q)=(x_i,p_j,q_k)$, where
\begin{eqnarray}
x_i=x_\mathrm{min}+\Delta x(i-1/2)&\mathrm{for}&1\leq i\leq N_x,\\
p_j=p_\mathrm{min}+\Delta p(j-1/2)&\mathrm{for}&1\leq j\leq N_p,\\
q_k=q_\mathrm{min}+\Delta q(k-1/2)&\mathrm{for}&1\leq k\leq N_q.
\end{eqnarray}
Next, we define the number of particles of a species $s$ in the cell at time $t$ as,
\begin{equation}
N^s_{ijk}(t)=\int^{x_i+\Delta x/2}_{x_i-\Delta x/2}dx
\int^{p_j+\Delta p/2}_{p_j-\Delta p/2}dp
\int^{q_k+\Delta q/2}_{q_k-\Delta q/2}dq f^s(x,p,q,t),
\label{5:number}
\end{equation}
and the energy of particles contained in the cell at $t$;
\begin{equation}
E^s_{ijk}(t)=\int^{x_i+\Delta x/2}_{x_i-\Delta x/2}dx
\int^{p_j+\Delta p/2}_{p_j-\Delta p/2}dp
\int^{q_k+\Delta q/2}_{q_k-\Delta q/2}dq \Gamma^sf^s(x,p,q,t).
\label{5:ene}
\end{equation}
On the other hand, we discretize electromagnetic fields by defining them only at the positions $x_i$,
\begin{equation}
E^\parallel_i(t)=E^\parallel(x_i,t),\ \ \ 
E^\perp_i(t)=E^\perp(x_i,t),\ \ \ 
B^\perp_i(t)=B^\perp(x_i,t).
\label{5:field}
\end{equation}

\subsection{\label{5:splitting}Splitting of equations}
Applying the operator splitting method, Equation (\ref{5:energy}) is numerically integrated by two steps. 
One is the step for the advection of particles and electromagnetic fields and the other is the step for the exchange of energy  between particles and electromagnetic fields. 

The Vlasov equation (\ref{5:vlasov}) is an advection equation with no source term, while the energy equation (\ref{5:energy}) contains advection terms and a source term. 
We split the energy equation (\ref{5:energy}) into the two parts as follows,
\begin{eqnarray}
&&\frac{\partial (\Gamma^sf^s)}{\partial t}+\frac{p}{\Gamma^s}\frac{\partial (\Gamma^s f^s)}{\partial x}
+R_q^s\left(E^\parallel+\frac{q}{\Gamma^s}B^\perp\right)\frac{\partial (\Gamma^sf^s)}{\partial p}
\nonumber\\
&&\hspace{10em}
+R_q^s\left(E^\perp-\frac{p}{\Gamma^s}B^\perp\right)\frac{\partial (\Gamma^sf^s)}{\partial q}=0,\label{5:energya}\\
&&\frac{\partial (\Gamma^sf^s)}{\partial t}=R_q^s\frac{pE^\parallel+qE^\perp}{\Gamma^s}f^s.\label{5:energyb}
\end{eqnarray}
One can see that the advection part of the energy equation (\ref{5:energya}) takes the same form as the Vlasov equation (\ref{5:vlasov}). 
Therefore, we introduce an operator ${\cal A}_p[E^\parallel(t),G(t),H(t),\Delta t]$ that evolves the variables $N_{ijk}^s(t)$ or $E_{ijk}^s(t)$ by a time interval $\Delta t$ according to Equations (\ref{5:vlasov}) or (\ref{5:energya}) for given $E^\parallel(t)$, $G(t)$, and $H(t)$ (or $E^\perp(t)$ and $B^\perp(t)$). 
For the interaction part, we introduce another operator  ${\cal I}_p[E^\parallel(t),G(t),H(t),\Delta t]$ that evolves the variable $E_{ijk}^s(t)$ by a time interval $\Delta t$ according to Equation (\ref{5:energyb}) with given $E^\parallel(t)$, $G(t)$, and $H(t)$.

We present a method to calculate the time evolution of the quantities defined by equations (\ref{5:number})-(\ref{5:field}).
As is the case for the energy equation, the Maxwell equations contain advection terms and source terms. 
Thus, we split them into the two parts as follows,
\begin{eqnarray}
&&\frac{\partial G}{\partial t}+\frac{\partial G}{\partial x}=0,\ \ \ \frac{\partial H}{\partial t}-\frac{\partial H}{\partial x}=0,\label{5:maxwella}\\
&&\frac{\partial E^\parallel}{\partial t}=-J^\parallel,\ \ \ \frac{\partial G}{\partial t}=-\frac{J^\perp}{2},\ \ \ 
\frac{\partial H}{\partial t}=-\frac{J^\perp}{2}.
\label{5:maxwellb}
\end{eqnarray}
Here we introduce two operators that evolve the variables $G_i(t)$ and $H_i(t)$ by a time interval $\Delta t$ according to Equations (\ref{5:maxwella}) as ${\cal A}_g[\Delta t]$ and ${\cal A}_h[\Delta t]$.
In addition, for the interaction part, we introduce three operators that evolve the variables $E^\parallel_i(t)$, $G_i(t)$, and $H_i(t)$ by a time interval $\Delta t$ according to Equation (\ref{5:maxwellb}) as ${\cal I}_e[\Delta t]$, ${\cal I}_g[\Delta t]$, and ${\cal I}_h[\Delta t]$. 
The explicit procedures of the thus introduced operators for advection terms are discussed in Sec. \ref{5:advection}. 
Sec. \ref{5:interaction} discusses those for source terms. 
Using the operators introduced above, we propose a scheme to numerically integrate the relativistic Vlasov-Maxwell system according to the following steps,
\begin{eqnarray}
\mathrm{step1}:&
N_{ijk}^{s*}&={\cal A}_p[E^\parallel(t),G(t),H(t),\Delta t/2]N_{ijk}^s(t)\nonumber\\
&E_{ijk}^{s*}&={\cal A}_p[E^\parallel(t),G(t),H(t),\Delta t/2]E_{ijk}^s(t)\nonumber\\
&G_i^*&={\cal A}_g[\Delta t]G_i(t)\nonumber\\
&H_i^*&={\cal A}_h[\Delta t]H_i(t)\\
&&\nonumber\\
\mathrm{step2}:&
E_{ijk}^{s**}&={\cal I}_p[E^\parallel(t),G^*,H^*,\Delta t]E_{ijk}^{s*}\nonumber\\
&E^{\parallel}_i(t+\Delta t)&={\cal I}_e[\Delta t]\nonumber\\
&G_i^{**}&={\cal I}_g[\Delta t]G_i^*\nonumber\\
&H_i^{**}&={\cal I}_h[\Delta t]H_i^*\\
&&\nonumber\\
\mathrm{step3}:&
N_{ijk}^{s}(t+\Delta t)&={\cal A}_p[E^{\parallel}_i(t+\Delta t),G^{**}_i,H^{**}_i,\Delta t/2]N_{ijk}^{s*}\nonumber\\
&E_{ijk}^{s}(t+\Delta t)&={\cal A}_p[E^{\parallel}_i(t+\Delta t),G^{**}_i,H^{**}_i,\Delta t/2]E_{ijk}^{s**}\nonumber\\
&G_i(t+\Delta t)&={\cal A}_g[\Delta t]G_i^{**}\nonumber\\
&H_i(t+\Delta t)&={\cal A}_h[\Delta t]H_i^{**}.
\end{eqnarray}
The electric current densities $J^\perp$ and $J^\parallel$, which are necessary for the integration of the source terms, are evaluated between step1 and step2. 
The procedure for the evaluation is explained in Sec. \ref{5:interpolation}.

\subsection{\label{5:advection}Advection part}
For the integration of the advection part, we make use of the characteristics of the Vlasov equation (\ref{5:vlasov}), \begin{equation}
\frac{dx}{dt}=\frac{p}{\Gamma^s},\ \ \ 
\frac{dp}{dt}=R_q^s\left(E^\parallel+\frac{q}{\Gamma^s}B^\perp\right),\ \ \ 
\frac{dq}{dt}=R_q^s\left(E^\perp-\frac{p}{\Gamma^s}B^\perp\right),
\label{5:characteristics}
\end{equation}
which are equivalent to the equation of motion of a relativistic charged particle, because there exists a reliable scheme for the integration of these equations widely used in PIC simulations \cite{BL91}, the Buneman-Boris method. 

\begin{figure}[b]
\begin{center}
\includegraphics[scale=0.5]{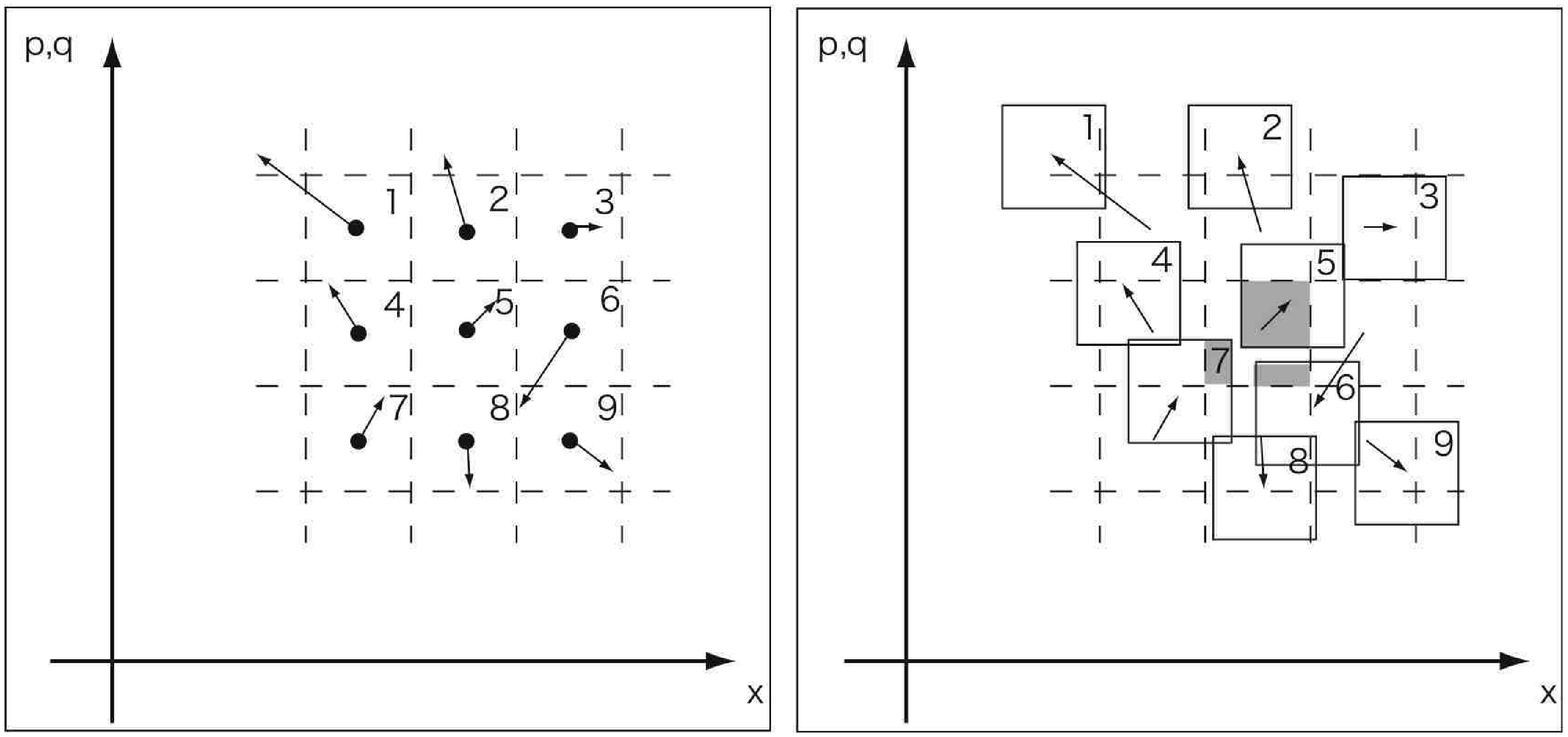}
\caption{Schematic views of the integration of the advection part.}
\label{figure1}
\end{center}
\end{figure}

At first, using the Buneman-Boris method, we obtain the orbit of a particle located at the center of each cell $(x_i,p_j,q_k)$ at time $t$. 
We thus calculate the coordinates $(x_i',p_j',q_k')$ of the particle at $t+\Delta t$ as 
\begin{eqnarray}
x_i'&=&x_i+\int^{t+\Delta t}_t\frac{p}{\Gamma^s}dt,\nonumber\\
p_j'&=&p_j+\int^{t+\Delta t}_tR_q^s\left(E^\parallel+\frac{q}{\Gamma^s}B^\perp\right)dt,\\\
q_k'&=&q_k+\int^{t+\Delta t}_tR_q^s\left(E^\perp-\frac{p}{\Gamma^s}B^\perp\right)dt.\nonumber
\end{eqnarray}
We then assume that the other particles  in the same cell move with the same velocity as the particle having been located at the center, which should be a good approximation for a sufficiently small cell. 
The relation between the size of the cell and the accuracy of the above treatment is discussed in \S3.7 and examined in \S4.1.
The intuitive explanation for the scheme is shown in Figure \ref{figure1}. 
In each panel, the horizontal axis represents the $x$-axis and the vertical axis represents the $p$- and $q$-axes. 
Although we draw the phase space as two dimensional, actual calculations are performed in the three dimensional phase space $(x,p,q)$. 
The procedure to calculate the number of particles in the cell (cell 5) located at the center of the surrounding nine cells at $t+\Delta t$ is as follows; 
(1) calculate the orbit of a particle located at the center of each cell (the left panel) using the Buneman-Boris method. 
(2) count the number of particles entering the original position of cell 5 under the assumption for a uniform distribution of particles inside each cell. 
In other words, the number of particles in cell 5 at $t+\Delta t$ is defined as that of particles located in the gray zones in the right panel of Figure \ref{figure1}. 
Therefore, an explicit expression of the operator ${\cal A}_p$ becomes
\begin{eqnarray}
{\cal A}_p[E^\parallel_i,G_i,H_i,\Delta t]N^s_{ijk}(t)&=&
\sum_{i'=i-1}^{i+1}\sum_{j'=j-1}^{j+1}\sum_{k'=k-1}^{k+1}
N^s_{i'j'k'}(t)\nonumber\\
&&\times\frac{|x_{i'}'-x_{i'}|}{\Delta x}\frac{|p_{j'}'-p_{j'}|}{\Delta p}\frac{|q_{k'}'-q_{k'}|}{\Delta q}.
\label{5:adv_p}
\end{eqnarray}
Here the summations in this expression run over only the cells overlapping the original position of the cell at $(x_i,p_j,q_k)$, i.e., cell 5, cell 6, and cell 7 for the case of Figure \ref{figure1}. 
We evolve the energy contained in a cell $E^s_{ijk}(t)$ in the same way. 
In this method, the number (or the mass) and  the kinetic energy of particles are conserved for each step. 

The advection part of the Maxwell equations (\ref{5:maxwella}) consists of two linear advection equations with a constant velocity that have exact solutions in the form of
\begin{equation}
G(x,t)=G(x-t,0),\ \ \ 
H(x,t)=H(x+t,0).
\end{equation}
Therefore, assuming $\Delta t=\Delta x$, one finds that the relations
\begin{equation}
G_i(t+\Delta t)=G_{i-1}(t),\ \ \ 
H_i(t+\Delta t)=H_{i+1}(t),
\label{5:adv_gh}
\end{equation}
hold. 
We use these relations for the integration of the advection part of the Maxwell equation. 
Because this method is based on the exact solution of a linear advection equation, no numerical diffusion occurs.

\subsection{\label{5:interpolation} Interpolation }
As we noted in Section \ref{5:splitting}, the electric current density needs to be evaluated for integration of the interaction part. 
In the following, we discuss a method to evaluate the electric current density. 
The key ingredient for the method is interpolation of the distribution function $f^s(x,p,q,t)$. 
We know the number $N_{ijk}^s(t)$ of particles and the energy $E_{ijk}^s(t)$ contained in each cell. 
From the definition of the two variables, the distribution function $f^s(x,p,q,t)$ must satisfy Equations (\ref{5:number}) and (\ref{5:ene}) for given $N_{ijk}^s(t)$ and $E_{ijk}^s(t)$. 
In other words, we have two constraints. 
So the interpolation function, which is defined as $f^s_{ijk}(t)$, generally have the form with two unknown coefficients, 
\begin{equation}
f^s_{ijk}(t)=a_{ijk}+b_{ijk}g(x,p,q),
\end{equation}
where $g(x,p,q)$ is a function and the coefficients $a_{ijk}$ and $b_{ijk}$ are determined from the constraints (\ref{5:number}) and (\ref{5:ene}). 
Here, to determine the form of the function $g(x,p,q)$, we consider the meaning of the constraints. 
The constraints, (\ref{5:number}) and (\ref{5:ene}), are the zero-order and the first-order moment of the Lorentz factor. 
Then, we assume the interpolation function $f^s_{ijk}(t)$ to take the form of
\begin{equation}
f^s_{ijk}(t)=a_{ijk}+b_{ijk}\Gamma^s,
\label{5:f_cell}
\end{equation}
We should note that there are many other candidates for the form of the interpolation function. 
If we calculate time evolutions of other macroscopic variable for each cell, e.g., momenta of particles, second-order moment of the Lorentz factor, and so on, or use the number $N_{i\pm 1j\pm 1,k\pm 1}^s(t)$ and the energy $E_{i\pm 1j\pm 1,k\pm 1}^s(t)$ of particles in neighboring cells, we can construct an interpolation function including more correction terms,
\begin{equation}
f^s_{ijk}(t)=a_{ijk}+b_{ijk}\Gamma^s+c_{ijk}h(x,p,q)+\cdots,
\end{equation}
where $h(x,p,q)$ is a function corresponding to the additional macroscopic  variable. 
In this study, we use the interpolation function (\ref{5:f_cell}), which is a linear function of the energy of particles, to reduce the computational cost. 

Substitution of the interpolation function (\ref{5:f_cell}) into the constraints and some algebraic manipulations lead to
\begin{eqnarray}
a_{ijk}&=&\frac{\langle(\Gamma^s)^2\rangle_{jk} N^s_{ijk}(t)-\langle\Gamma^s\rangle_{jk} E^s_{ijk}(t)}
{\Delta x\Delta p\Delta q[\langle(\Gamma^s)^2\rangle_{jk}-\langle\Gamma^s\rangle_{jk}^2]},\\
b_{ijk}&=&\frac{E^s_{ijk}(t)-\langle\Gamma^s\rangle_{jk} N^s_{ijk}(t)}
{\Delta x\Delta p\Delta q[\langle(\Gamma^s)^2\rangle_{jk}-\langle\Gamma^s\rangle_{jk}^2]},
\end{eqnarray}
where the bracket represents the following integral,
\begin{equation}
\langle A\rangle_{jk}=\frac{1}{\Delta p\Delta q}\int^{p_j+\Delta p/2}_{p_j-\Delta p/2}\int^{q_k+\Delta q/2}_{q_k-\Delta q/2}Adpdq.
\end{equation}
Appendix \ref{appendixD} gives the expressions of the variables $\langle\Gamma^s\rangle_{jk}$ and $\langle(\Gamma^s)^2\rangle_{jk}$. 
Thus the distribution function $f^s_{ijk}$ takes a uniform value in each spatial cell $i$. 
Using this interpolation function, the electric current densities due to a particle species $s$ are evaluated as
\begin{eqnarray}
j_{ijk}^{s\parallel}&=&R^s_q
\int ^{p_j+\Delta p/2}_{p_j-\Delta p/2}\int ^{q_k+\Delta q/2}_{q_k-\Delta q/2}
\frac{p}{\Gamma^s}f^s_{ijk}(t)dpdq\nonumber\\
&=&R^s_q\Delta p\Delta q\left(a_{ijk}\langle \frac{p}{\Gamma^s}\rangle_{jk}+b_{ijk}\langle p\rangle_{jk}\right)
\\
j_{ijk}^{s\perp}&=&R^s_q
\int ^{p_j+\Delta p/2}_{p_j-\Delta p/2}\int ^{q_k+\Delta q/2}_{q_k-\Delta q/2}
\frac{q}{\Gamma^s}f^s_{ijk}(t)dpdq\nonumber\\
&=&R^s_q\Delta p\Delta q\left(a_{ijk}\langle \frac{q}{\Gamma^s}\rangle_{jk}+b_{ijk}\langle q\rangle_{jk}\right)
\label{5:int_current}
\end{eqnarray}
in non-dimensional forms. 

However, the thus constructed function $f^s_{ijk}$ is not guaranteed to take positive values at all points in the region $[p_j-\Delta p/2,p_j+\Delta p/2]\times[q_k-\Delta q/2,q_k+\Delta q/2]$ for each $i$. 
Because the distribution function of real plasmas must be positive at any point in the phase space,  the interpolation is modified if $f^s_{ijk}$ takes a negative value. 
We use the following simple expressions for $j^{s\parallel}_{ijk}$ and $j^{s\perp}_{ijk}$,
\begin{eqnarray}
j_{ijk}^{s\parallel}&=&
R^s_q\langle\frac{p}{\Gamma^s}\rangle \frac{N^s_{ijk}(t)}{\Delta x},
\\
j_{ijk}^{s\perp}&=&
R^s_q\langle\frac{q}{\Gamma^s}\rangle \frac{N^s_{ijk}(t)}{\Delta x},
\end{eqnarray}
instead of the expressions (\ref{5:int_current}) in cells with negative $f^s_{ijk}(t)$.

One can evaluate the electric current density by summing up these variables as
\begin{equation}
J^\parallel_i=\sum_s\sum_{j}\sum_{k}j^{s\parallel}_{ijk},\ \ \ 
J^\perp_i=\sum_s\sum_{j}\sum_{k}j^{s\perp}_{ijk} .
\label{5:relj}
\end{equation}

\subsection{\label{5:interaction}Interaction part}
In this subsection, we propose a method to integrate the interaction part with respect to time. 
This method conserves the sum of the kinetic energy of particles and the electromagnetic energy. 

Equations (\ref{5:maxwellb}) are discretized as
\begin{eqnarray}
E^\parallel_i(t+\Delta t)&=&E^\parallel_i(t)-J^\parallel(t+\Delta t/2)\Delta t,\nonumber\\
G_i(t+\Delta t)&=&G_i(t)-\frac{J^\perp(t+\Delta t/2)}{2}\Delta t,\label{egh}\\
H_i(t+\Delta t)&=&H_i(t)-\frac{J^\perp(t+\Delta t/2)}{2}\Delta t,\nonumber
\end{eqnarray}
where the electric current densities are evaluated beforehand according to the procedure described in the previous subsection. 
These equations give expressions for the operators ${\cal I}_e$, ${\cal I}_g$, and ${\cal I}_h$. 
On the other hand, to obtain the energy $E^s_{ijk}(t)$ of particles in a cell evolved by the interaction part of the energy equation (\ref{5:energyb}), Equation (\ref{5:energyb}) is integrated with respect to $p$ and $q$ as
\begin{eqnarray}
&&\frac{\partial }{\partial t}\int^{p_j+\Delta p/2}_{p_j-\Delta p/2}\int^{q_k+\Delta q/2}_{q_k-\Delta q/2}
\Gamma^s f^s dpdq\nonumber\\
&&\hspace{10em}=E^\parallel\int^{p_j+\Delta p/2}_{p_j-\Delta p/2}\int^{q_k+\Delta q/2}_{q_k-\Delta q/2}\frac{p}{\Gamma^s}f^sdpdq\\
&&\hspace{11em}+E^\perp\int^{p_j+\Delta p/2}_{p_j-\Delta p/2}\int^{q_k+\Delta q/2}_{q_k-\Delta q/2}\frac{q}{\Gamma^s}f^sdpdq,\nonumber
\end{eqnarray}
which means that the total energy of particles in a cell is changed by interactions between particles and electric fields. 
Discretizing this equation, we then propose the following scheme for the integration:
\begin{equation}
\frac{E^s_{ijk}(t+\Delta t)-E^s_{ijk}(t)}{\Delta t}=
\frac{E^\parallel(t+\Delta t)+E^\parallel(t)}{2}j^{s\parallel}_{ijk}
+\frac{E^\perp(t+\Delta t)+E^\perp(t)}{2}j^{s\perp}_{ijk},
\label{5:Ip}
\end{equation}
which gives an expression for the operator ${\cal I}_p$. 

In the following, we will show that this procedure conserves the total energy. 
Summing up the above equation with respect to $j$, $k$, and $s$, and then substituting the relations (\ref{5:relj}) and (\ref{egh}) into the result, one obtains
\begin{eqnarray}
\sum_{jks}E_{ijk}^s(t+\Delta t)-\sum_{jks}E_{ijk}^s(t)&=&
\left[E^\parallel_i(t)-\frac{J^\parallel_i(t)}{2}\Delta t\right]J^\parallel_i(t)\Delta t\label{5:energy1}\\
&&+\left[G_i(t)+H_i(t)-\frac{J^\perp_i(t)}{2}\Delta t\right]J^\perp_i(t)\Delta t,\nonumber
\end{eqnarray}
which represents the change of kinetic energy of particles after a time step in this scheme. 
The change of the electromagnetic energy is obtained by summing the square of each of (\ref{egh}),
\begin{eqnarray}
&&\frac{[E^\parallel_i(t+\Delta t)]^2}{2}+[G_i(t+\Delta t)]^2+[H_i(t+\Delta t)]^2\nonumber\\
&&\hspace{2em}=
\frac{[E^\parallel_i(t)]^2}{2}+[G_i(t)]^2+[H_i(t)]^2
-\left[E^\parallel_i(t)-\frac{J^\parallel_i(t)}{2}\Delta t\right]J^\parallel_i(t)\Delta t\nonumber\\
&&\hspace{4em}
-\left[G_i(t)+H_i(t)-\frac{J^\perp_i(t)}{2}\Delta t\right]J^\perp_i(t)\Delta t
\label{5:energy2}
\end{eqnarray}
By summing up both sides of Equations (\ref{5:energy1}) and (\ref{5:energy2}) with respect to $i$, one can easily check that the total energy in a region $[x_i-\Delta x/2,x_i+\Delta x/2]\times[p_\mathrm{min},p_\mathrm{max}]\times[q_\mathrm{min},q_\mathrm{max}]$ is conserved:
\begin{equation}
\sum_{ijks}E_{ijk}^s(t)+
\frac{[E^\parallel_i(t)]^2}{2}+[G_i(t)]^2+[H_i(t)]^2=\mathrm{const}.
\end{equation}
In other words, the procedures ${\cal I}_e$, ${\cal I}_g$, ${\cal I}_h$, and ${\cal I}_p$ expressed by (\ref{egh}) and (\ref{5:Ip}) give a conservative scheme for the integration of the interaction part of the relativistic Vlasov-Maxwell system.

\subsection{Conditions for the time interval}
In Sections \ref{5:splitting}, \ref{5:advection}, \ref{5:interpolation}, and \ref{5:interaction}, we present procedures that evolve the number and the energy of particles in a cell and electromagnetic fields. 
In order for the procedures to work, the time interval $\Delta t$ is required to satisfy some conditions.  

As we noted in Section \ref{5:advection}, the scheme (\ref{5:adv_gh}) requires that the time interval $\Delta t$ must be equal to $\Delta x$. 
Furthermore, the scheme for integration of the advection part of the Vlasov equation (\ref{5:adv_p}) requires that the displacement of a particle by integration of Equations (\ref{5:characteristics}) along the $x$-, $p$-, and $q$-axes must not exceed the intervals $\Delta x$, $\Delta p$, and $\Delta q$. 
In short, particles must not jump over a cell. 
These conditions impose the value of the time interval to satisfy
\begin{equation}
\Delta t=\Delta x,\ \ \ 
\Delta t< \frac{\Delta p}{\mathrm{max}(|E^\parallel_i|+|B^\perp_i|)},\ \ \ 
\Delta t< \frac{\Delta q}{\mathrm{max}(|E^\perp_i|+|B^\perp_i|)},
\end{equation}
where $\mathrm{max}(A_i)$ represents the maximum of the variable $A_i$ for all $i$.

\subsection{Accuracy of the scheme\label{accuracy}}
Finally, we mention the accuracy of our scheme proposed in this section. 
As explained above, our scheme is based on various procedures, such as splitting of equations, the advection part, the interaction part, and the evaluation of the current density, which makes the mathematical proof of the accuracy of our scheme very difficult. 
Then, we estimate the accuracy of the advection of particles, which is likely to be the most inaccurate compared to the other procedure. 

In the procedure solving the advection part of the Vlasov equation, all particles in a given cell in the phase space are assumed to move with the same orbit as that of the particle located at the center of the cell. 
However, this treatment obviously involves errors to a certain extent, because particles located at the different position from the center must be integrated under different initial conditions. 
In particular, the difference is most significant for particles located at the vertex of the cell. 
Since the difference of the position between particles at the vertex and the center is of the order of $\Delta x,\Delta p$, and $\Delta q$, the estimated positions of particles at the vertex contain errors of the order of $\Delta x,\Delta p$, and $\Delta q$, which indicates the number of particles in the cell at the next step contains errors of the order of $\Delta x,\Delta p$, and $\Delta q$. 
Therefore, the procedure solving the advection part of the Vlasov equation have first order accuracy in the physical and the momentum spaces. 

\section{Test problems\label{5:test}}

In this section, we show results of simulations performed by using the scheme proposed in the previous section. 
For the purpose, we solve three test problems, the gyration of particles, the Weibel instability, and the wakefield acceleration. 
The gyration of particles is solved to investigate into the accuracy of our scheme. 
The Weibel instability and the wakefield acceleration are well-known plasma processes and important in both experimental and astrophysical contexts. 

\subsection{Gyration of particles}
We assume that electrons are uniformly distributed in the physical space with a gaussian distribution in the momentum spaces, 
\begin{equation}
f^e(x,p,q,0)\propto \exp\left(-\frac{p^2+q^2}{\sigma^2}\right),
\end{equation} 
where $\sigma$ represents the dispersion in the momentum spaces, and an uniform magnetic field,
\begin{equation}
E^\parallel(x,0)=E^\perp(x,0)=0,\ \ \ 
B^\perp(x,0)=B_0, 
\end{equation}
where $B_0$ is a constant. 
One can easily check that the above configuration is a stationary solution of the Vlasov-Maxwell system. 
However, since our scheme suffers from a numerical diffusion as expected in the previous section, the distribution function $f^e(x,p,q,t)$ at $t$ must be different slightly from the initial one $f^e(x,p,q,0)$. 
Then, we adopt the following value $\epsilon$ as a measure of the accuracy of our scheme, 
\begin{equation}
\epsilon=\sqrt{\frac{\sum_{ijk}[N^e_{ijk}(t_g)-N^e_{ijk}(0)]^2}{N_xN_pN_q}},
\end{equation}
where $t_\mathrm{g}$ represent the gyration period given by $m_\mathrm{e}c/(eB_0)$. 
Figure \ref{accuracy} shows the result with $\sigma^2=2.0$ and $B_0=1.0$. 
The ranges of the space coordinate, the longitudinal momentum, and the lateral momentum are given by $x\in [-\sqrt{2\pi},\sqrt{2\pi}]$, $p,q\in [-10,10]$, respectively. 
The periodic boundary is imposed in the physical space, while, in the momentum space, the free boundary condition is imposed. 
The filled circles represent the values $\epsilon$ for various $N_x$($=110, 120, 130, 140, 150, 160, 170, 180, \mathrm{and\ } 190$) and fixed $N_p$ and $N_q$ ($N_p=N_q=200$), whereas the filled squares represent those for $N_p=110, 120, 130, 140, 150, 160, 170, 180,$ and $190$ and $N_x=N_q=200$. 

The solid line shows that the value $\epsilon$ seems to scale roughly as $(\Delta x)^{1.8}$. 
The value $\epsilon$ is expected to depend strongly on $\Delta p$ and $\Delta q$ rather than $\Delta x$ in this test problem where particles rotate in the momentum space $(p,q)$. 
In other words, the dependence of $\epsilon$ on $\Delta x$  have  uncertainty because of the insensitiveness. 
Therefore, we conclude that the dependence derived above must be $\epsilon\propto (\Delta x)^2$ essentially. 
However, this does not mean second order accuracy in the physical space. 
Because of the condition $\Delta t=\Delta x$ mentioned in \S 3.6, when we double the number of zones $N_x$ in the physical space, the time interval $\Delta t$ must be half of the previous value. 
Therefore, the value $\epsilon$ scales as $\Delta t\Delta x$, which indicates that our scheme has first order accuracy in time and the physical space. 
On the other hand, the dashed line shows that the value $\epsilon$ scales as $\Delta p$, which confirms the estimation in \S 3.7.

\subsection{Weibel instability}
The Weibel instability is a kind of plasma instabilities caused by anisotropic momentum distributions of collisionless plasma. 
The formulation and dispersion relation of the Weibel instability operating in a relativistic one-dimensional plasma are shown in Appendix \ref{appendixE}. 

For a simulation of the Weibel instability, we treat ions as a uniform background and assume that electrons have the following initial distribution,
\begin{equation}
f^e(x,p,q,0)=\delta (p)\frac{\delta (q-q_\mathrm{b})+\delta (q+q_\mathrm{b})}{2},
\label{5:initial}
\end{equation}
which is approximated in the discretized form by
\begin{equation}
N_{ijk}^e(0)=\left\{
\begin{array}{ccl}
1/2&\mathrm{for}&-\Delta p/2<p_j<\Delta p/2\\
&&\ \ \ \ \mathrm{and}\ q_\mathrm{b}-\Delta q/2<q_k<q_\mathrm{b}+\Delta q/2,\\
1/2&\mathrm{for}&-\Delta p/2<p_j<\Delta p/2\\ 
&&\ \ \ \ \mathrm{and}\ -q_\mathrm{b}-\Delta q/2<q_k<-q_\mathrm{b}+\Delta q/2,\\
0&\mathrm{otherwise}&
\end{array}\right.
\label{initialN}
\end{equation}
\begin{equation}
E_{ijk}^e(0)=\left\{
\begin{array}{ccl}
\sqrt{1+q_\mathrm{b}^2}/2&\mathrm{for}&-\Delta p/2<p_j<\Delta p/2\\ 
&&\ \ \ \ \mathrm{and}\ q_\mathrm{b}-\Delta q/2<q_k<q_\mathrm{b}+\Delta q/2,\\
\sqrt{1+q_\mathrm{b}^2}/2&\mathrm{for}&-\Delta p/2<p_j<\Delta p/2\\
&&\ \ \ \ \mathrm{and}\ -q_\mathrm{b}-\Delta q/2<q_k<-q_\mathrm{b}+\Delta q/2,\\
0&\mathrm{otherwise}&
\end{array}\right.
\label{initialE}
\end{equation}
Here we have introduced a constant $q_\mathrm{b}$ that represents the bulk momentum of the counter-stream of the plasma. 
The initial configuration of the electromagnetic field is
\begin{equation}
E^\parallel=E^\perp=0,\ \ \ B^\perp=\epsilon\cos(kx),
\end{equation}
where $\epsilon$ is a small parameter ($=10^{-5}$) and $k$ is the wave number of the perturbation. 

We calculate the evolution of a plasma with the above initial condition in the simulation domain whose spatial interval is given by $x\in[-\pi/k,\pi/k]$.
The longitudinal momentum ranges are given by $p\in[-5,5]$ for $q_\mathrm{b}=2.065$ (the corresponding bulk velocity is $0.9c$), $p\in[10,-10]$ for $q_\mathrm{b}=7.018$ (0.99c), and $p\in[30,-30]$ for $q_\mathrm{b}=22.344$ (0.999c). 
The lateral momentum range is $q\in[-5,5]$. 
The periodic boundary is imposed in the $x$ direction, while, in the momentum space, the free boundary condition is imposed. 

Figure \ref{figure2} shows the time evolutions of the kinetic energy $K_\mathrm{e}$ of electrons, the electric energy $E$, and the magnetic energy $B$ defined by
\begin{eqnarray}
K_\mathrm{e}&=&\sum_{ijk}E^e_{ijk}(t),\\
E&=&\frac{\Delta x}{2}\sum_{i}[(E^\parallel_i)^2+(E^\perp_i)^2]\nonumber\\&=&
\frac{\Delta x}{2}\sum_{i}[(E^\parallel_i)^2+(G_i+H_i)^2],\\
B&=&\frac{\Delta x}{2}\sum_{i}(B^\perp_i)^2
=\frac{\Delta x}{2}\sum_{i}(G_i-H_i)^2,
\end{eqnarray}
for the case of $q_\mathrm{b}=2.065$ (the corresponding bulk velocity is $0.9c$) and $k=1$. 
The numbers of zones for the three coordinates are $N_x=100$, $N_p=N_q=50$ for $q_\mathrm{b}=2.065$, $N_x=100$, $N_p=100$, $N_q=50$ for $q_\mathrm{b}=7.018$, and $N_x=100$, $N_p=300$, $N_q=50$ for $q_\mathrm{b}=22.344$. 
The dashed line in Figure \ref{figure2} reproduces the growth rate calculated from the linearized analysis with $P_\mathrm{th}=0.1$ described in Appendix \ref{appendixE}. 
Although we treat a cold plasma whose initial momentum distribution is given by (\ref{5:initial}), the initial setup (\ref{initialN}) has particles with  some dispersions in the momenta of the order of the width of the momentum bins $\Delta p=\Delta q=0.2$. 
Therefore we compare the dispersion relation from the numerical simulations with that derived from linearized analyses with a finite temperature corresponding to the size of the momentum bin. 
The numerical simulation seems to reproduce the theoretical growth rate for a given wave number $k=1$.

Figure \ref{figure3} summarizes the growth rates for other cases as a function of the wave number of the perturbation. 
The lines in this figure represent the growth rates for the bulk velocities $0.9c$, $0.99c$, and $0.999c$ calculated from the dispersion relation (\ref{5:dispersion}) with $P_\mathrm{th}=0.1$. 
The plotted points show values measured from results of the simulation. 


\subsection{Wakefield acceleration}
The wakefield acceleration is a promising mechanism for the acceleration of particles to highly relativistic speeds (see, \citet{e96}, for review). 
The ponderomotive force of a coherent electromagnetic wave propagating in a stationary plasma, such as an intense laser in laboratory or a light pulse emitted by a certain active phenomenon in astrophysical environment \cite{ctt02,l06,h08},  excites a longitudinal electric field and efficiently generates high-energy particles. 

To simulate such circumstances, we impose the following boundary condition for the electromagnetic field,
\begin{equation}
G(0,t)=A_0\omega_\mathrm{L}\exp\left[-\frac{(t-2\tau)^2}{\tau^2}\right]\sin(\omega_\mathrm{L}t),
\ \ \ H(0,t)=0,
\label{5:boundary}
\end{equation}
which produces a linearly polarized electromagnetic wave (light pulse) propagating in the $+x$-direction. 
Here we have introduced some parameters characterizing simulations; $A_0$ the scale of the vector potential, $\omega_\mathrm{L}$ the frequency of the light pulse, $\tau$ the duration of the light pulse. 
For the initial configuration of particles, we consider a cold, homogeneous, stationary plasma composed of electrons.
The momentum distribution is expressed as
\begin{equation}
 f^e(x,p,q,0)=\delta(p)\delta(q),
\end{equation}
which leads to
\begin{equation}
N_{ijk}^e(0)=E_{ijk}^e(0)=\left\{
\begin{array}{ccl}
1&\mathrm{for}&-\Delta p/2<p_j<\Delta p/2\\
&&\ \ \ \ \mathrm{and}\ -\Delta q/2<q_k<\Delta q/2,\\
0&\mathrm{otherwise}&
\end{array}\right.
\label{5:ini_wake}
\end{equation}
We treat ions as a neutralizing background, choosing the value of the scale of the light pulse to be longer than the electron inertial length $c/\omega_\mathrm{e}$ but shorter than the ion inertial length $\sqrt{m_\mathrm{i}/m_\mathrm{e}}c/\omega_\mathrm{e}$. 

Some results for the case of $A_0=2.0$, $\omega_\mathrm{L}=2.0$, and $\tau=\pi/2$ are shown in Figures \ref{figure5}, \ref{figure6}, and \ref{figure7}. 
In each figure, the panels represent the color-coded $q$-integrated distribution function, the longitudinal electric field, the transverse electric field, and the transverse magnetic field from top to bottom. 
It is clearly seen that a sinusoidal electrostatic field is excited immediately after the passage of the light pulse and accelerate electrons, resulting in some bunches of electrons in the phase space. 
The responses of the plasma and the electric field to the light pulse is consistent with the previous studies. 
\citet{s90} studied this process by numerically solving equations which treat non-linear interactions of particles and waves (see also, \citet{t90}). 
They showed that sawtooth-like longitudinal waves associated with some bunches of particles in the phase space form after the passage of a light pulse. 
Recent two-dimensional PIC simulations (see, e.g., \citet{kskth08}) show a similar behavior. 
The behavior is also reproduced in our results, which indicates that the method presented here can treat the correct behavior of the distribution function of relativistic, collisionless plasmas. 

The electron distributions in the momentum space at $t=200$ and $x=178,184$ are plotted in Figure \ref{figure8}. 
It is clearly seen that the existence of high-energy electrons up to $p=17m_\mathrm{e}c$ (the corresponding velocity is equal to 0.998c) at $x=178$, where the strong electrostatic field, i.e., the wakefield, is excited due to the ponderomotive force. 
At $x=184$, on the contrary, there exists no accelerated electron, since the node of the wakefield is located at this point.

\section{Discussion and Conclusions\label{5:conclusion}}
In this paper, we have proposed a new conservative scheme for numerical integration of the relativistic Vlasov-Maxwell system and performed three test problems, the gyration of particles, the Weibel instability, and the wakefield acceleration. 
Adopting semi-Lagrange method, we succeed in developing a scheme that conserves the number of particles and the sum of the energies of particles and electromagnetic fields. 
Since the previous scheme \cite{blgsb08} solving the relativistic Vlasov-Maxwell system do not treat the dispersion of the lateral momentum of particles, our scheme is the first one that can treat the dispersion correctly. 
Results of the simulations clearly indicate that our method succeeds in reproducing detailed behaviors of the distribution functions in the phase space. 
Especially, the tail of the distributions where only a tiny fraction of particles reside seems to be solved with considerably high accuracy, while PIC simulations would suffer from large statistical error there. 

As we noted above, Vlasov simulations generally require more computational resources than PIC simulations do. 
Furthermore, as previous works \cite{mcct02,cmvv07} have investigated, Vlasov simulations suffer from so-called "filamentation problem". 
Ref.\cite{cmvv07} studied wave-particle interactions of a plasma approaching to an equilibrium state using PIC simulation and showed that the equilibrium is realized through a phase mixing accompanied by formation of filamentary structures in the phase space. 
In Vlasov description, particles composing a plasma are treated as a continuous medium, which means that Vlasov equation can not take account of essential discreteness of plasma. 
As a result, artificial entropy may arise when a structure with the characteristic scale smaller than the mesh size is generated in the phase space. 
Ref.\cite{cmvv07} argued that this artificial entropy prevents the plasma treated by the Vlasov simulation from following the correct path toward the statistical equilibrium.
Therefore, we need careful studies of long-term evolutions of plasmas by using Vlasov simulations. 

Nevertheless, Vlasov simulations provide us detailed dynamics of plasmas in the phase space. 
Though we deal with the 1D2V relativistic Vlasov-Maxwell system, our method can be applied to the 2D3V and 3D3V cases
Although our scheme proposed in this paper suffer from numerical diffusion, there is a plenty room for improvement. 
For example, in order to integrate advection part of the Vlasov equation, we can use the orbit of the particle located at each vertex of a cell. 
In other words, taking account for the deformation of the cell at each time step improves the accuracy of the scheme. 
In theoretical investigations into complex behaviors of collisionless plasmas, Vlasov simulations must be a attractive tool to compensate defects of PIC simulations.

\section{Acknowledgments}
The authors are grateful to the anonymous referees for their constructive comments on this manuscript. 
This work was supported by Grant-in-Aid for JSPS Fellows 21$\cdot$1726. 

\appendix
\section{\label{appendixD}Evaluation of Some Integrals} 
In this section, we evaluate some integrals used to construct the interpolation function in Sec.3.4. 
The first one is the average of the modified Lorentz factor (\ref{mLorentz}) in the phase space, $\langle \Gamma^s\rangle_{jk}$, defined by
\begin{equation}
\langle \Gamma^s\rangle_{jk}=\frac{1}{\Delta p\Delta q}\int^{p_j+\Delta p/2}_{p_j-\Delta p/2}
\int^{q_k+\Delta q/2}_{q_k-\Delta q/2}\sqrt{(R^s_m)^2+p^2+q^2}dpdq.
\end{equation}
To perform the integrations, we introduce a function of the variables $p$ and $q$ described by
\begin{eqnarray}
g_1(p,q)&=&\frac{pq}{3}\sqrt{(R^s_m)^2+p^2+q^2}
-\frac{(R^s_m)^3}{3}\mathrm{Arctan}\left[\frac{pq}{R^s_m\sqrt{(R^s_m)^2+p^2+q^2}}\right]\nonumber\\
&&+\frac{p}{6}[p^2+3(R^s_m)^2]\ln \left[q+\sqrt{(R^s_m)^2+p^2+q^2}\right]\nonumber\\
&&+\frac{q}{6}[q^2+3(R^s_m)^2]\ln \left[p+\sqrt{(R^s_m)^2+p^2+q^2}\right].
\end{eqnarray}
Since the differentiation with respect to $p$ and the subsequent differentiation with respect to $q$ of this function leads to
\begin{equation}
\frac{\partial ^2g_1}{\partial p\partial q}=\sqrt{(R^s_m)^2+p^2+q^2},
\end{equation}
one can evaluate the integral $\langle \Gamma^s\rangle$ as
\begin{eqnarray}
\langle \Gamma^s\rangle_{jk}&=&
\frac{g_1(p_j+\Delta p/2,q_k+\Delta q/2)}{\Delta p\Delta q}
-\frac{g_1(p_j+\Delta p/2,q_k-\Delta q/2)}{\Delta p\Delta q}\nonumber\\
&&-\frac{g_1(p_j-\Delta p/2,q_k+\Delta q/2)}{\Delta p\Delta q}
+\frac{g_1(p_j-\Delta p/2,q_k-\Delta q/2)}{\Delta p\Delta q}. 
\end{eqnarray}
The second integral is the average of the square of the modified Lorentz factor in the phase space, $\langle (\Gamma^s)^2\rangle_{jk}$, defined by
\begin{equation}
\langle (\Gamma^s)^2\rangle_{jk}=\frac{1}{\Delta p\Delta q}\int^{p_j+\Delta p/2}_{p_j-\Delta p/2}
\int^{q_k+\Delta q/2}_{q_k-\Delta q/2}\left[(R^s_m)^2+p^2+q^2\right]dpdq.
\end{equation}
This integration is straightforward and one obtains
\begin{equation}
\langle (\Gamma^s)^2\rangle_{jk}=
(R^s_m)^2+p_j^2+\frac{\Delta p^2}{12}+q_k^2+\frac{\Delta q^2}{12}.
\end{equation}
The third and forth integrals are defined by
\begin{equation}
\bigg\langle \frac{p}{\Gamma^s}\bigg\rangle_{jk}=
\frac{1}{\Delta p\Delta q}\int^{p_j+\Delta p/2}_{p_j-\Delta p/2}
\int^{q_k+\Delta q/2}_{q_k-\Delta q/2}\frac{p}{\sqrt{(R^s_m)^2+p^2+q^2}}dpdq,
\end{equation}
and
\begin{equation}
\bigg\langle \frac{q}{\Gamma^s}\bigg\rangle_{jk}=
\frac{1}{\Delta p\Delta q}\int^{p_j+\Delta p/2}_{p_j-\Delta p/2}
\int^{q_k+\Delta q/2}_{q_k-\Delta q/2}\frac{q}{\sqrt{(R^s_m)^2+p^2+q^2}}dpdq,
\end{equation}
respectively. 
To evaluate the third integral, we define the following function,
\begin{equation}
g_2(p,q)=\frac{q}{2}\sqrt{(R^s_m)^2+p^2+q^2}+
\frac{(R^s_m)^2+p^2}{2}\ln\left[q+\sqrt{(R^s_m)^2+p^2+q^2}\right].
\end{equation}
Since the differentiation with respect to $q$ and the subsequent differentiation with respect to $p$ leads to
\begin{equation}
\frac{\partial}{\partial p}\left(\frac{\partial g_2}{\partial q}\right)=
\frac{p}{\sqrt{(R^s_m)^2+p^2+q^2}},
\end{equation}
the third integral is written as
\begin{eqnarray}
\bigg\langle \frac{p}{\Gamma^s}\bigg\rangle_{jk}&=&
\frac{g_2(p_j+\Delta p/2,q_k+\Delta q/2)}{\Delta p\Delta q}
-\frac{g_2(p_j+\Delta p/2,q_k-\Delta q/2)}{\Delta p\Delta q}\nonumber\\
&&-\frac{g_2(p_j-\Delta p/2,q_k+\Delta q/2)}{\Delta p\Delta q}
+\frac{g_2(p_j-\Delta p/2,q_k-\Delta q/2)}{\Delta p\Delta q}. 
\end{eqnarray}
Using the same function, the forth integral is expressed as
\begin{eqnarray}
\bigg\langle \frac{q}{\Gamma^s}\bigg\rangle_{jk}&=&
\frac{g_2(q_k+\Delta q/2,p_j+\Delta p/2)}{\Delta p\Delta q}
-\frac{g_2(q_k-\Delta q/2,p_j+\Delta p/2)}{\Delta p\Delta q}\nonumber\\
&&-\frac{g_2(q_k+\Delta q/2,p_j-\Delta p/2)}{\Delta p\Delta q}
+\frac{g_2(q_k-\Delta q/2,p_j-\Delta p/2)}{\Delta p\Delta q}. 
\end{eqnarray}
The remaining integrals $\langle p\rangle_{jk}$ and $\langle q\rangle_{jk}$ can be evaluated by a straightforward manner:
\begin{equation}
\langle p\rangle_{jk}=p_j,\ \ \ 
\langle q\rangle_{jk}=q_k.
\end{equation}

\section{\label{appendixE}The Relativistic Weibel Instability}
The dispersion relation of the Weibel instability in both nonrelativistic and relativistic plasmas have already been derived in several investigations (see, e.g., \citet{cpbm98}). 
Nevertheless, we review the formulation and the dispersion relation of the Weibel instability in a relativistic one-dimensional plasma for completeness of this paper. 

\subsection{Formulation}
We assume that the initial state characterized by an unperturbed distribution function $f_0^s$ has no electromagnetic field and that the space ($x$) and time ($t$) dependences of the perturbations are proportional to $\exp[i(kx-\omega t)]$. 
We then consider how the perturbation $\delta f^s$ on the distribution function and the lateral components of the electromagnetic fields $\delta E^\perp$ and $\delta B^\perp$ evolve according to the relativistic Vlasov-Maxwell system. 
The linearized relativistic Vlasov equation expressed as
\begin{equation}
\left(-i\omega+ik\frac{p}{\Gamma^s}\right)\delta f^s+R^s_q\frac{q}{\Gamma^s}\delta B^\perp
\frac{\partial f_0^s}{\partial p}+R_q^s\left(\delta E^\perp-\frac{p}{\Gamma^s}\delta B^\perp\right)\frac{\partial f_0^s}{\partial q}=0,
\label{5:lv}
\end{equation}
and the linearized Maxwell equations 
\begin{eqnarray}
-i\omega \delta E^\perp+ik\delta B^\perp&=&-\delta J^\perp,\label{5:lma} \nonumber\\
-i\omega \delta B^\perp+ik\delta E^\perp&=&0\label{5:lmb}
\end{eqnarray}
govern the time evolutions of the perturbed quantities. 
Elimination of $\delta B^\perp$ in Equations (\ref{5:lv}) by using Equation (\ref{5:lmb}) yields
\begin{equation}
\left(-i\omega+ik\frac{p}{\Gamma^s}\right)\delta f^s+R^s_q\frac{q}{\Gamma^s}\frac{k}{\omega}\delta E^\perp
\frac{\partial f_0^s}{\partial p}+R_q^s\left(1-\frac{p}{\Gamma^s}\frac{\omega}{k}\right)\delta E^\perp\frac{\partial f_0^s}{\partial q}=0
\end{equation}
and the expression for $\delta f^s$,
\begin{equation}
\delta f^s=-\frac{iR^s_q}{\omega}\left(\frac{qk}{\omega\Gamma^s-kp}\frac{\partial f^s_0}{\partial p}+
\frac{\partial f^s_0}{\partial q}\right)\delta E^\perp.
\end{equation}
The perturbed electric current density $\delta J^\perp$ is related to the perturbed distribution function $\delta f^s$ as
\begin{equation}
\delta J^\perp=\sum_sR^s_q\int^\infty_{-\infty}dp\int^\infty_{-\infty}dq\frac{q}{\Gamma^s}\delta f^s,
\end{equation}
and necessary to obtain the dispersion relation.
We then assume that ions are uniformly distributed in the physical space with no bulk velocity,
\begin{equation}
f_0^i=n_0\delta (p)\delta (q),
\end{equation}
where $\delta (x)$ represents the delta function, and that electrons have the following form of the initial distribution,
\begin{eqnarray}
f_0^e&=&n_0\frac{\Theta(p+P_\mathrm{th})-\Theta(p-P_\mathrm{th})}{2P_\mathrm{th}}
\\
&&\times\frac{\Theta(q+P_0+P_\mathrm{th})-\Theta(q+P_0-P_\mathrm{th})+
\Theta(q-P_0+P_\mathrm{th})-\Theta(q-P_0-P_\mathrm{th})}{4P_\mathrm{th}}\nonumber,
\end{eqnarray}
where $\Theta(x)$ represents the Heaviside function. 
The parameter $P_\mathrm{th}$ represents the thermal dispersion of the momentum distribution of electrons, and $P_0$ represents their bulk momentum. 
This assumption means that only electrons contribute to the generation of the electric current density. 
We define the following two integrals,
\begin{eqnarray}
I_1&=&\int^\infty_{-\infty}dp\int^\infty_{-\infty}dq\frac{q}{\Gamma^e}
\frac{qk}{\omega\Gamma^e-kp}\frac{\partial f^e_0}{\partial p},\\
I_2&=&\int^\infty_{-\infty}dp\int^\infty_{-\infty}dq\frac{q}{\Gamma^e}
\frac{\partial f^e_0}{\partial q},
\end{eqnarray}
which contribute to the electric current density $\delta J^\perp$, and evaluate them beforehand. 

From the properties of the Heaviside function, the first integral reduces to
\begin{equation}
I_1=-\frac{n_0k^2}{2P_\mathrm{th}}\int^{P_0+P_\mathrm{th}}_{P_0-P_\mathrm{th}}
\frac{q^2dq}{\sqrt{1+P_\mathrm{th}^2+q^2}\left[\omega^2(1+P_\mathrm{th}^2+q^2)-k^2P_\mathrm{th}^2\right]}.
\end{equation}
We then define the function $g(q)$ in the form of
\begin{eqnarray}
g(q)=&&\sqrt{\omega^2(1+P_\mathrm{th}^2)-k^2P_\mathrm{th}^2}\mathrm{Arctan}
\left[\frac{kP_\mathrm{th}}{\sqrt{\omega^2(1+P_\mathrm{th}^2)-k^2P_\mathrm{th}^2}}
\frac{q}{\sqrt{1+P_\mathrm{th}^2+q^2}}\right]\nonumber\\
&&-kP_\mathrm{th}\log\left(q+\sqrt{1+P_\mathrm{th}^2+q^2}\right)
\end{eqnarray}
Since the derivative of this function is
\begin{equation}
\frac{dg}{dq}=
-\frac{\omega^2kP_\mathrm{th}}{\sqrt{1+P_\mathrm{th}^2+q^2}}
\frac{q^2}{\omega^2(1+P_\mathrm{th}^2+q^2)-k^2P_\mathrm{th}^2},
\end{equation}
one can express the integral $I_1$ as
\begin{equation}
I_1=\frac{n_0k}{2P_\mathrm{th}^2\omega^2}[g(P_0+P_\mathrm{th})-g(P_0-P_\mathrm{th})].
\end{equation}
On the other hand, the second integral becomes
\begin{eqnarray}
I_2&=&-\frac{n_0}{2P_\mathrm{th}^2}
\left\{(P_0+P_\mathrm{th})\mathrm{Arcsinh}\left[\frac{P_\mathrm{th}}{\sqrt{1+(P_0+P_\mathrm{th})^2}}\right]\right.
\nonumber\\
&&\hspace{5em}\left.-(P_0-P_\mathrm{th})\mathrm{Arcsinh}\left[\frac{P_\mathrm{th}}{\sqrt{1+(P_0-P_\mathrm{th})^2}}\right]
\right\}
\end{eqnarray}
from the straightforward evaluation of the integral. 

Using these integrals, the perturbed electric current density is expressed as
\begin{equation}
\delta J^\perp=-\frac{i}{\omega}(I_1+I_2)\delta E^\perp.
\end{equation}
Substitution of this expression into Equation (\ref{5:lma}) and non-trivial $\delta B^\perp$ yields the following dispersion relation,
\begin{equation}
\omega^2-k^2+\frac{\omega_\mathrm{p}}{n_0}(I_1+I_2)=0
\label{5:dispersion}
\end{equation}

\subsection{Dispersion relation}
\subsubsection{cold plasmas}
Before we solve the dispersion relation (\ref{5:dispersion}) with a fixed wave number $k$ and obtain the frequency $\omega$, we simplify Equation (\ref{5:dispersion}) by taking cold limit ($P_\mathrm{th}\rightarrow 0$) to clarify whether any unstable mode exists or not. 
Under the cold limit, the integrals defined in the previous section becomes
\begin{equation}
\lim_{P_\mathrm{th}\rightarrow 0}I_1=
-\frac{n_0k^2}{\omega^2}\frac{P_0^2}{(1+P_0^2)^{3/2}},
\end{equation}
and
\begin{equation}
\lim_{P_\mathrm{th}\rightarrow 0}I_2=
-\frac{n_0}{(1+P_0^2)^{3/2}},
\end{equation}
respectively. 
Then, the dispersion relation becomes
\begin{equation}
\omega^4-\left[k^2+\frac{1}{(1+P_0^2)^{3/2}}\right]\omega^2-\frac{k^2P_0^2}{(1+P_0^2)^{3/2}}=0.
\label{5:Dcold}
\end{equation}
One of the solution of this equation is a pure imaginary number, which means that this dispersion relation contains at least one unstable mode. 
The growth rates of the mode $\gamma$ defined by $i\gamma=\omega$ versus wave number $k$ for initial bulk velocities $0.9c$, $0.99c$, and $0.999c$ are plotted as solid lines in Figures \ref{figure9}, \ref{figure10}, and \ref{figure11}. 

\
\subsubsection{warm plasmas}
The analysis performed in the previous subsection indicates that there exists an unstable mode. 
Solving the dispersion relation of warm plasmas (\ref{5:dispersion}), one obtains the growth rate of the Weibel instability for plasmas with finite temperature. 
The results for the case of $P_\mathrm{th}=0.1$ and the bulk velocities $0.9c$, $0.99c$, and $0.999c$ are also shown in Figures \ref{figure9}, \ref{figure10}, and \ref{figure11}. 
One can see that the growth of unstable modes is suppressed in the high wave number regime. 

\begin{figure}
\begin{center}
\includegraphics[scale=0.5]{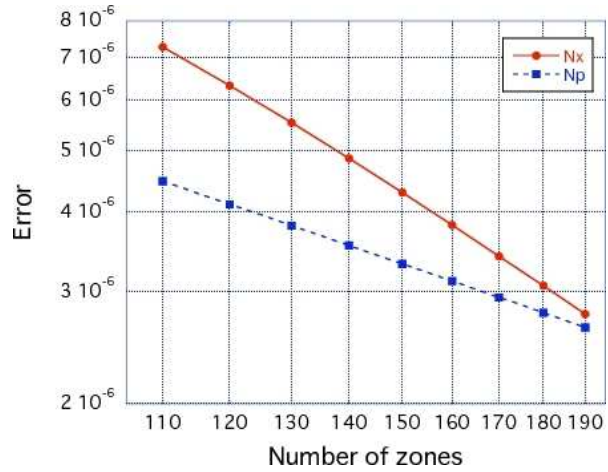}
\caption{Errors as a function of the number of zones.}
\label{accuracy}
\end{center}
\end{figure}
\begin{figure}
\begin{center}
\includegraphics[scale=0.6]{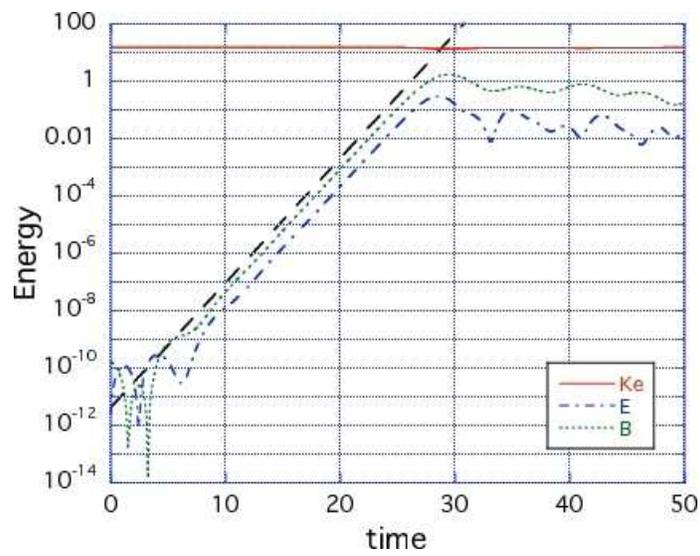}
\caption{The time evolution of the electron energy, $K_\mathrm{e}$ (solid), the electric energy $E$ (dash-dotted), and the magnetic energy $B$ (dashed). 
The dashed line represents the theoretical growth rate derived by the linearized analysis.}
\label{figure2}
\end{center}
\end{figure}
\begin{figure}
\begin{center}
\includegraphics[scale=0.6]{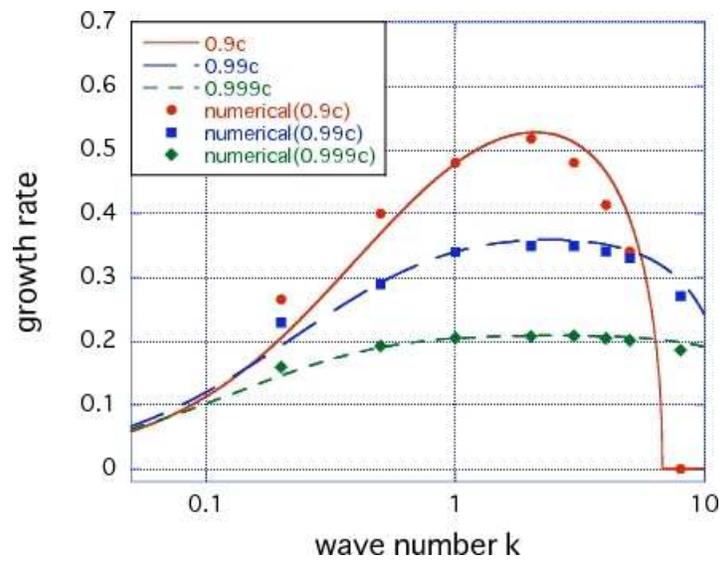}
\caption{The dispersion relation of the Weibel instability. 
The horizontal axis represents the wave number of perturbation and the vertical axis represents the corresponding growth rate. 
The solid, dashed, and dotted lines corresponds to the case that the bulk velocity is $0.9c$, $0.99c$, and $0.999c$. 
The points plotted the plane is the value measured from results of the simulation. }
\label{figure3}
\end{center}
\end{figure}

\begin{figure}
\begin{center}
\includegraphics[scale=0.6]{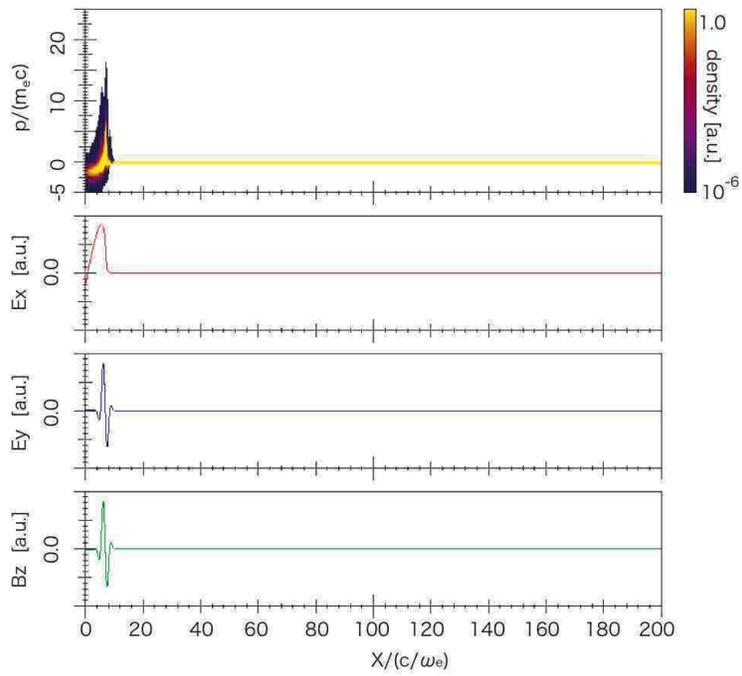}
\caption{Snapshot of the distribution function and the electromagnetic fields at $t=10$.}
\label{figure5}
\end{center}
\end{figure}
\begin{figure}
\begin{center}
\includegraphics[scale=0.6]{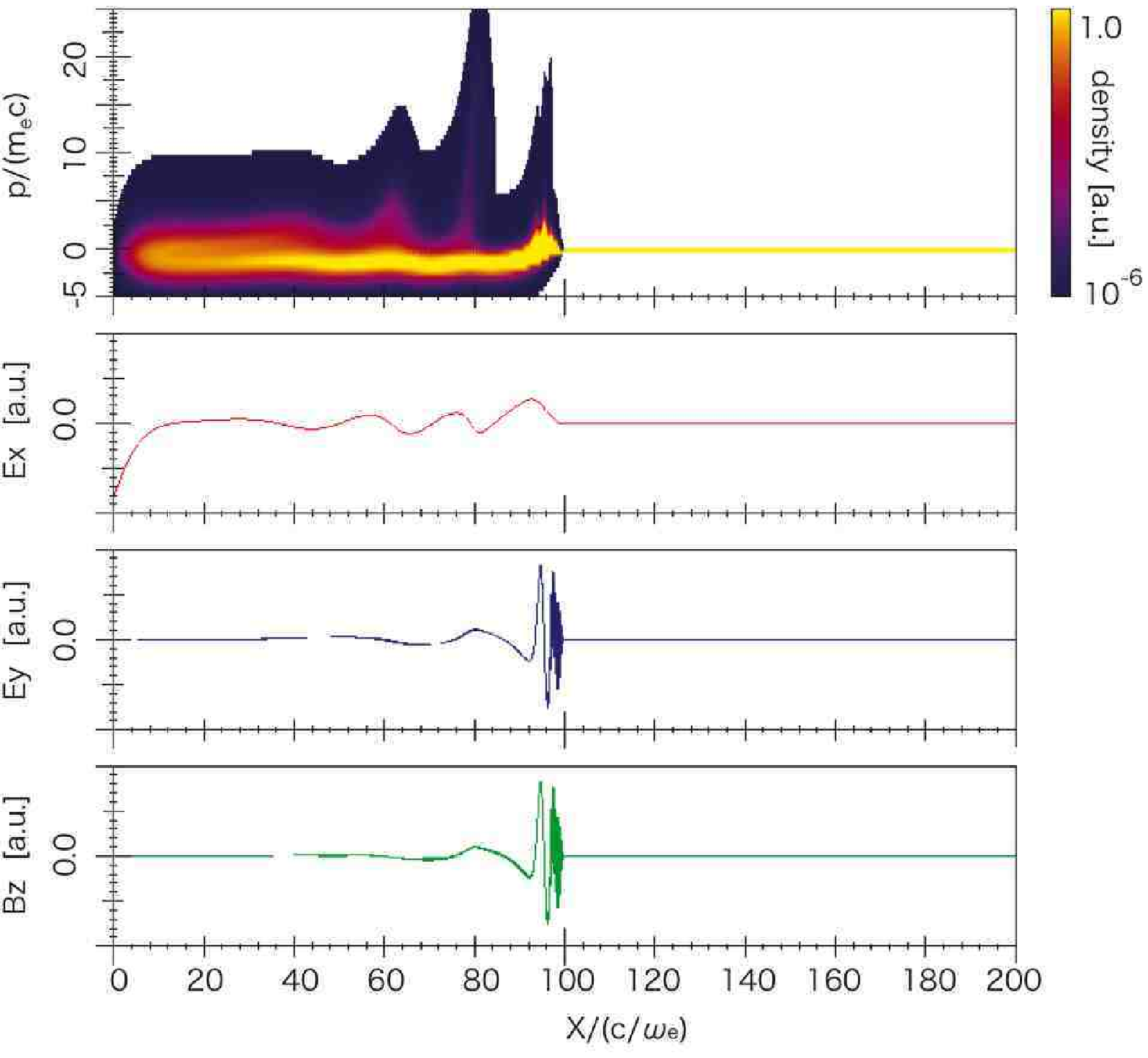}
\caption{Same as Figure \ref{figure5}, but for $t=100$.}
\label{figure6}
\end{center}
\end{figure}
\begin{figure}
\begin{center}
\includegraphics[scale=0.6]{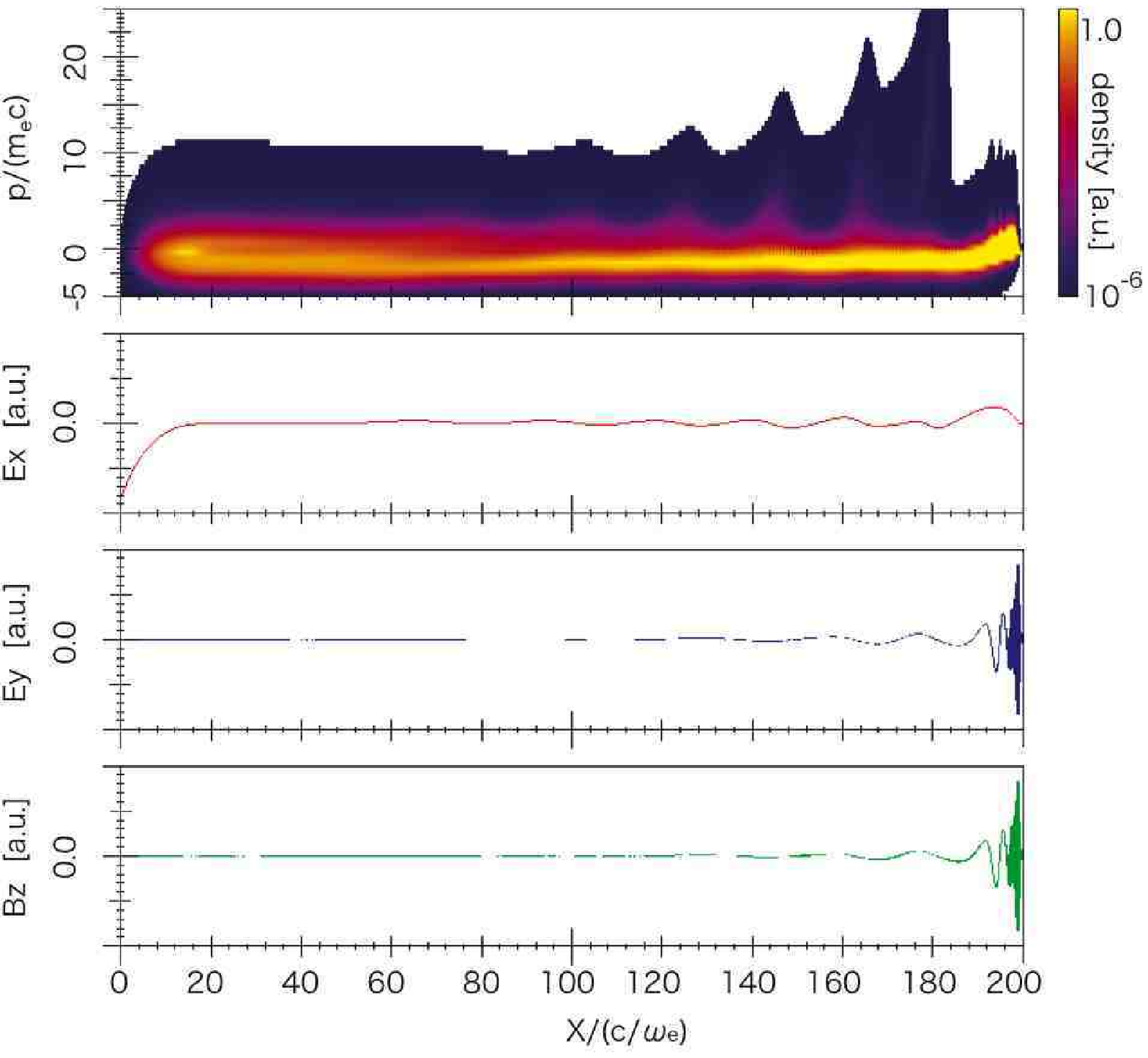}
\caption{Same as Figure \ref{figure5}, but for $t=200$.}
\label{figure7}
\end{center}
\end{figure}
\begin{figure}
\begin{center}
\includegraphics[scale=0.6]{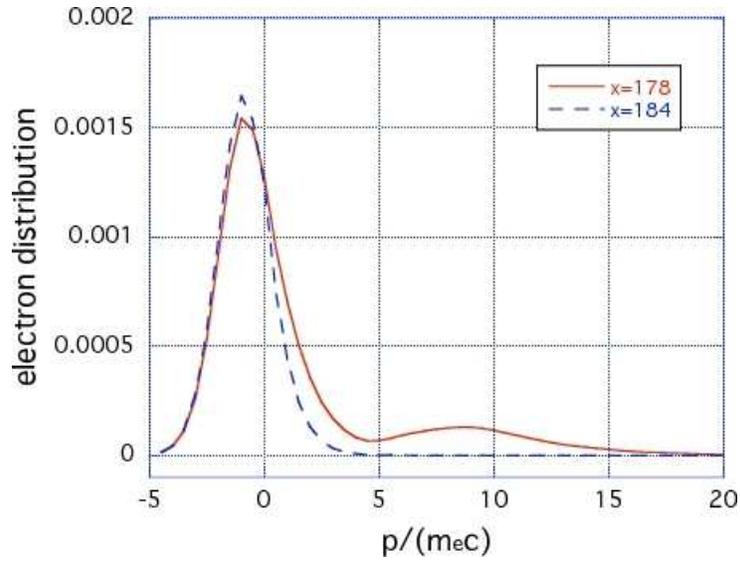}
\caption{The $q$-integrated electron distribution in the longitudinal momentum space at $t=200$ and $x=178,184$.}
\label{figure8}
\end{center}
\end{figure}
\begin{figure}
\begin{center}
\includegraphics[scale=0.6]{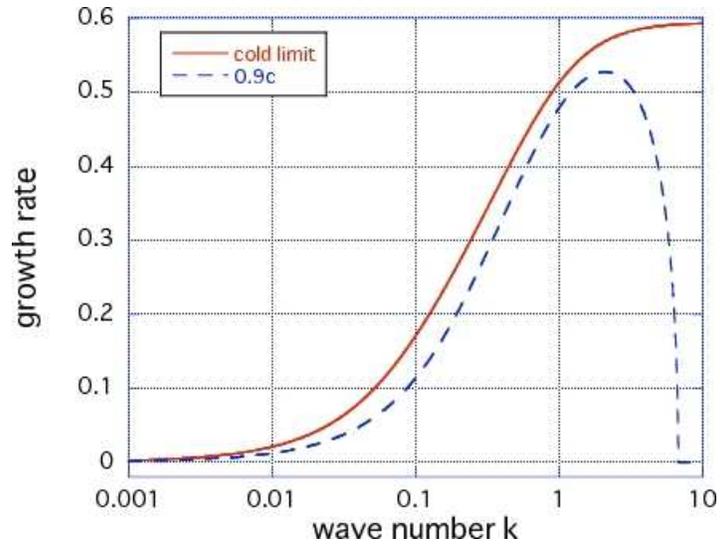}
\caption{The growth rate of the relativistic Weibel instability with the bulk velocity $0.9c$ as functions of wave numbers. 
The solid line represents the growth rate calculated from the dispersion relation for cold plasmas (\ref{5:Dcold}). 
The dashed line represents the growth rate calculated from the dispersion relation for warm plasmas with $P_\mathrm{th}=0.1$ (\ref{5:dispersion}).}
\label{figure9}
\end{center}
\end{figure}
 \begin{figure}
\begin{center}
\includegraphics[scale=0.6]{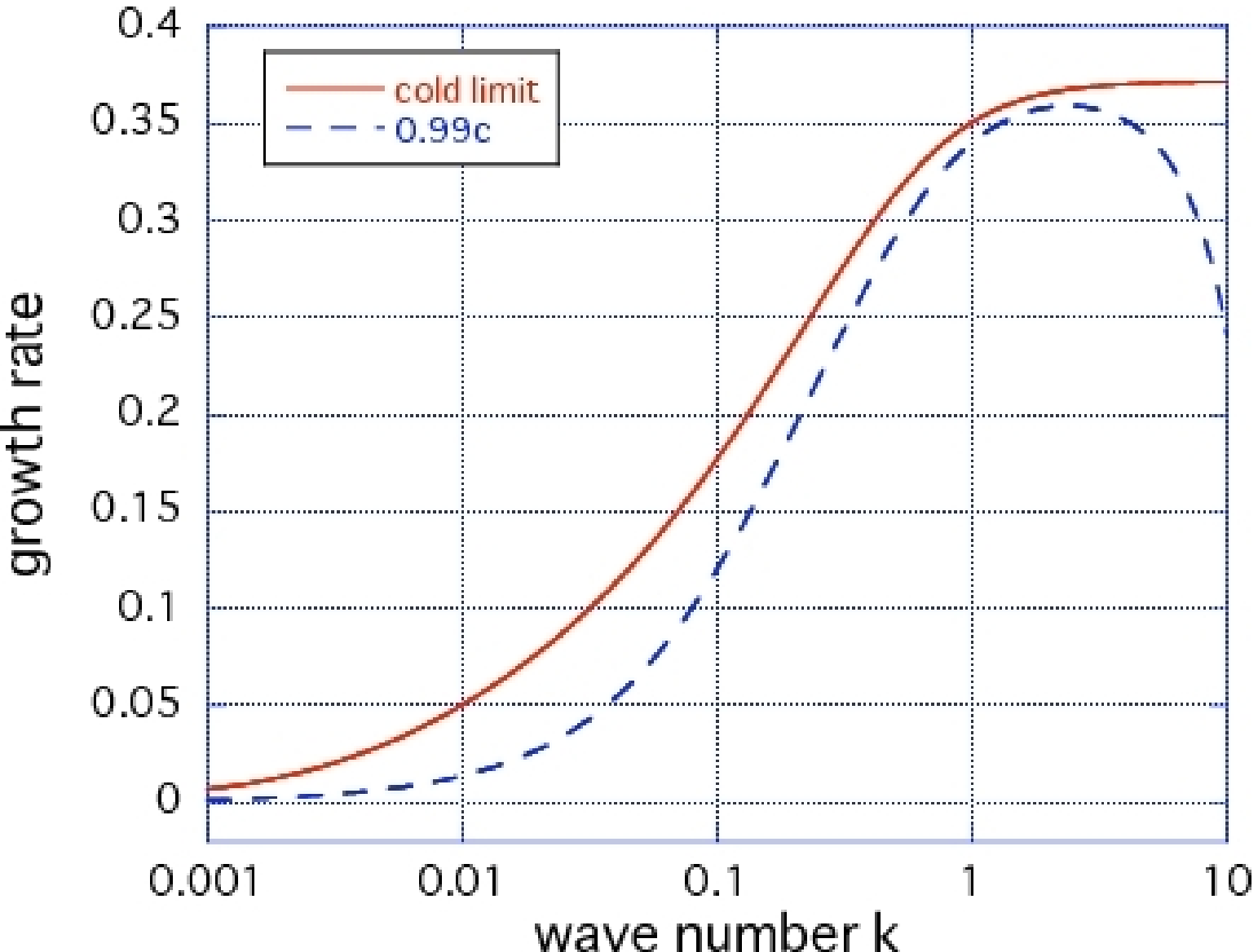}
\caption{Same as Figure \ref{figure9}, but for the bulk velocity $0.99c$.}
\label{figure10}
\end{center}
\end{figure}
\begin{figure}
\begin{center}
\includegraphics[scale=0.6]{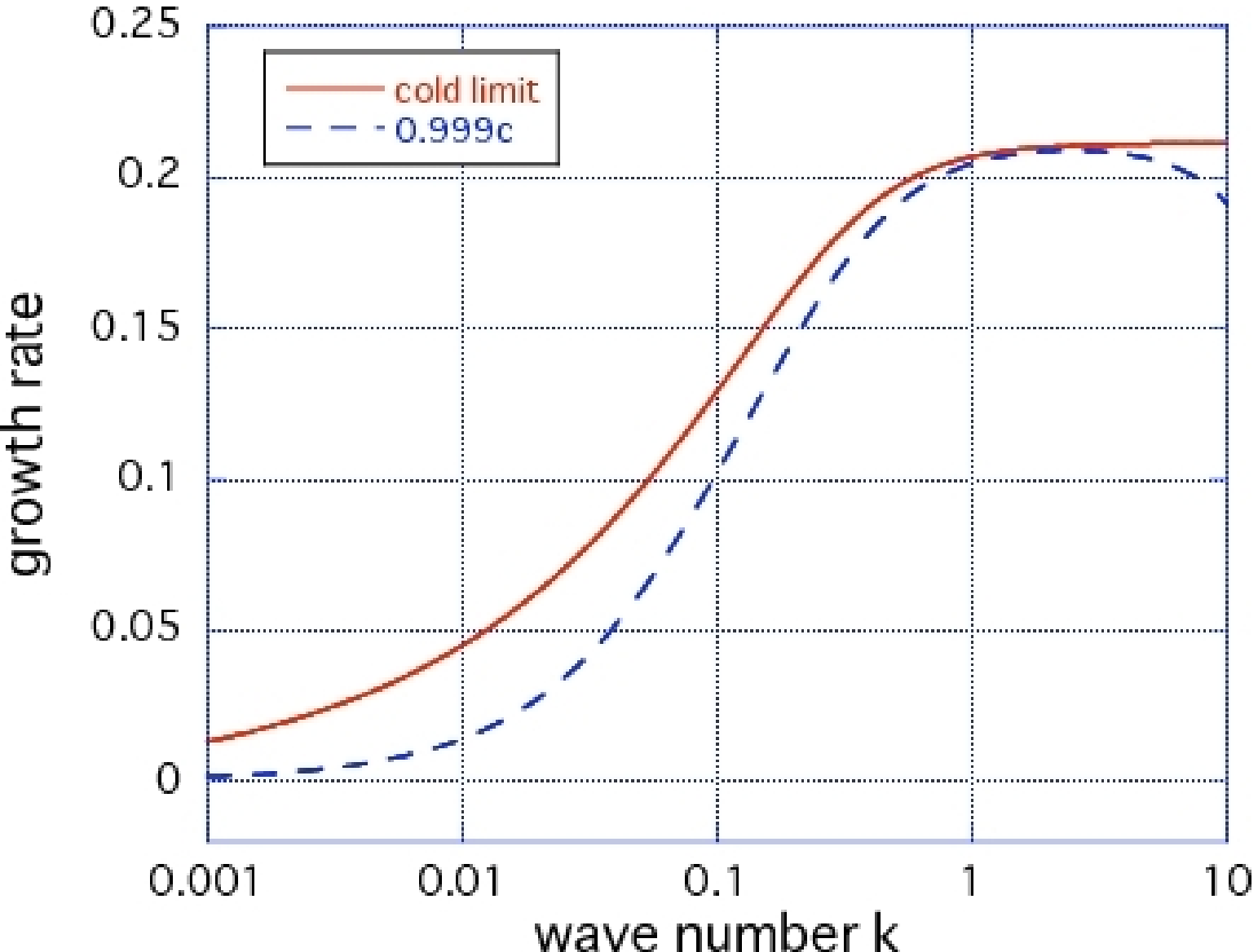}
\caption{Same as Figure \ref{figure9}, but for the bulk velocity $0.999c$.}
\label{figure11}
\end{center}
\end{figure}

\end{document}